\documentclass[11pt]{article}

\newlength{\vshift}
\newlength{\hshift}
\usepackage{amssymb,amsopn}

\def\nn{\nonumber }

\def\be{\beta}
\def\a{\alpha}

\def\g{\gamma}

\def\ds{\stackrel{\star}{,}}

\def\tr{{\rm Tr}}

\def\htR{\hat {R}}
\def\htom{\hat {\omega}}
\def\htE{\hat {E}}

\def\nn{\nonumber}
\def\be{\begin{equation}}             \def\ee{\end{equation}}
\def\ba#1{\begin{array}{#1}}          \def\ea{\end{array}}
\def\bea{\begin{eqnarray} }           \def\eea{\end{eqnarray} }
\def\beann{\begin{eqnarray*} }        \def\eeann{\end{eqnarray*} }
\def\beal{\begin{eqalign}}            \def\eeal{\end{eqalign}}
             
\def\bsubeq{\begin{subequations}}     \def\esubeq{\end{subequations}}
\def\bitem{\begin{itemize}}           \def\eitem{\end{itemize}}
                    
\def\pa{\partial}
\def\a{\alpha}
\def\b{\beta}

\def\d{\delta}
\def\e{\epsilon}                
\def\g{\gamma}

\def\k{\kappa}
\def\l{\lambda}
\def\m{\mu}
\def\n{\nu}

\def\r{\rho}                    
\def\s{\sigma}                  

\def\L{\Lambda}

\begin{document}

\begin{titlepage}

$\,$

\vspace{1.5cm}
\begin{center}

{\LARGE{\bf AdS-inspired noncommutative gravity on the Moyal plane}}

\vspace*{1.3cm}

{{\bf Marija Dimitrijevi\' c$^1$,
Voja Radovanovi\' c$^1$  and Hrvoje \v Stefan\v ci\'c$^2$}}

\vspace*{1cm}

$^1$University of Belgrade, Faculty of Physics\\
Studentski trg 12, 11000 Beograd, Serbia \\[1em]

$^2$Rudjer Bo\v skovi\' c Institute, Theoretical Physics Division\\
Bijeni\v cka 54, 10002 Zagreb, Croatia\\

\end{center}

\vspace*{2cm}

\begin{abstract}

We consider noncommutative gravity on a space with canonical noncommutativity that is 
based on the commutative MacDowell-Mansouri action. Gravity is treated as gauge theory of the noncommutative 
$SO(1,3)_\star$ group and the
Seiberg-Witten (SW) map is used to express noncommutative fields in terms of the
corresponding commutative fields. In the commutative limit the noncommutative action
reduces to the Einstein-Hilbert action plus the cosmological term and the topological
Gauss-Bonnet term. After the SW expansion in the noncommutative parameter the first
order correction to the action, as expected, vanishes.
We calculate the second order correction and write it in a manifestly
gauge covariant way.

\end{abstract}
\vspace*{1cm}



\vspace*{1cm}
\quad\scriptsize{eMail:
dmarija,rvoja@ipb.ac.rs; shrvoje@thphys.irb.hr}
\vfill

\end{titlepage}\vskip.2cm

\newpage
\setcounter{page}{1}
\newcommand{\Section}[1]{\setcounter{equation}{0}\section{#1}}
\renewcommand{\theequation}{\arabic{section}.\arabic{equation}}

\section{Introduction}

Field theories on noncommutative (NC) spaces have been investigated in many aspects
during the last twenty years. Various approaches to definition and analysis of
the properties of noncommutative spaces are present in the literature \cite{NCknjige}. One
of the the most frequently used is the approach of deformation quantization
\cite{DefQuant}. In this approach
noncommutative functions  $\hat{f}(\hat{x})$ are mapped to the functions of
commuting coordinates $f(x)$ and the abstract algebra multiplication is represented by
$\star$-product, which is a deformation of the usual point-wise multiplication. The
simplest and the most analyzed example of the $\star$-product is the  Moyal-Weyl
$\star$-product
\begin{equation}
\label{moyal}  f (x)\star  g (x) =
      e^{\frac{i}{2}\,\theta^{\m\n}\frac{\pa}{\pa x^\m}\frac{\pa}{ \pa
      y^\n}} f (x) g (y)|_{y\to x}\ ,
\end{equation}
defined by a constant antisymmetric matrix $\theta^{\mu\nu}$. Using this type of deformation
various problems were investigated: NC scalar field theories, NC gauge theories,
deformations of supersymmetric theories,\dots.

An important boost to the formulation of NC gauge theories came with the paper of Seiberg
and Witten \cite{Seiberg:1999vs}. They found a connection between gauge
theories on the commutative space and the corresponding NC gauge theories. This result 
was then used
by Wess et al. \cite{jssw} to formulate the enveloping algebra approach to NC
gauge theories. Namely, for groups which are of importance for physical
applications like $SU(N)$, NC gauge transformations close in the enveloping algebra.
This implies that the NC gauge field is also enveloping algebra-valued which leads to
(infinitely many) new degrees of freedom. However, using the Seiberg-Witten
(SW) map one can express all enveloping algebra-valued NC variables (gauge parameter and
fields) in terms of the corresponding commutative variables. In that way both theories
have the same number of degrees of freedom. This approach enabled the analysis of
renormalizability of NC gauge theories  \cite{Buric:2005xe}, anomalies \cite{martin},
construction of a NC deformation of the Standard Model \cite{cjsww} and investigation of its
phenomenological consequences \cite{tramp}.

On the other hand, construction of a NC generalization of General Relativity (GR) proved 
to be a difficult task. One of the reasons for this is the underlying symmetry of GR,
the diffeomorphism symmetry. One can follow the twist approach in which the commutative
diffeormorphisms are replaced by the twisted diffeomprphisms
\cite{defgt}. However, a full understanding of the twisted symmetries is still
missing. Having in mind that the SW approach works very well for NC gauge theories,
many authors consider NC gravity as a gauge theory of the Lorentz/Poincar\' e group.
This can be done in the Einstein-Cartan formalism \cite{MilutinKnjiga}. In this formalism
the gauge field for the local Lorentz symmetry, the spin connection
$\omega= \omega_\m {\rm d}x^\m$ and the vielbein $e=e_\m{\rm d}x^\m$  are independent
fields. The action is given by
\begin{equation}
S_{EC} = \frac{1}{16\pi G_N}\int {\rm d}^4 x \epsilon_{abcd}\epsilon^{\mu\nu\rho\sigma}
R^{\ \ ab}_{\mu\nu} e^{\ c}_\rho e^{\ d}_\sigma , \label{ECaction}
\end{equation}
with the curvature tensor $R_{\mu\nu} = \partial_\mu\omega_\nu - \partial_\nu\omega_\mu
-i [\omega_\mu, \omega_\nu]$. Indices
$a,b,\dots$ are local Lorentz (or flat) indices, while indices
$\mu,\nu,\dots$ are Einstein or space-time indices. The condition that the torsion
vanishes, $T^a_{\mu\nu} =0$ follows from the equations of motion. Using this condition
one can express the spin connection in terms of the vielbeins and the theory reduces to the
General Relativity with the metric $g_{\m\n} = \eta_{ab} e_\m^{\ a}e_\n^{\ b}$. The local
Lorentz symmetry can be generalized to the NC local Lorentz symmetry in different ways.
In \cite{ChPL} a deformation of the Einstein-Hilbert action is proposed. The construction
is based on the Wigner-Inon\"u contraction of the noncommutative $SO(1,4)_\star$ gauge
symmetry. However, noncommutative correction to the gravity action was not found.
In \cite{ChPrd04} the starting point was $GL(2,C)_\star$ NC gauge symmetry. This theory in
the commutative limit gives a gravity theory with complex metric. In \cite{dZanon03}
additional conditions (like vanishing torsion in the lowest order of the NC parameter) are
used to break the $U(2,2)_\star$ symmetry to the $SO(1,3)_\star$. Some authors formulate a
deformation of gravity theory on NC spaces with space-time dependent noncommutativity
\cite{BMS-07,MXZ-11}. Coupling of the NC gravity described by (\ref{ECaction}) with
fermions and gauge fields was discussed in \cite{PC09} and \cite{PC12}. It was shown there
that if reality of the NC gravity action is imposed, all odd order corrections
(in the NC parameter) have to vanish. Actually, the vanishing of the first order correction 
has already been shown in \cite{ChPL, dZanon03, Mukherjee2}. The first non-vanishing correction is the second
order correction. So far, no complete result for the second order correction was found,
mainly because the calculations are too involved.

In this paper we follow the approach of \cite{stelle-west}. Our starting point is the local
$SO(2,3)$ symmetry on four dimensional Minkowski space-time. The group $SO(2,3)$ has
$10$ generators, leading to the $10$ gauge fields in the theory. After the symmetry
breaking, the symmetry reduces to the $SO(1,3)$ and the generators split into $6$ fields
which correspond to the spin connection and $4$ vielbeins. The gravity action obtained after
 spontaneous symmetry breaking is identical to the action obtained by MacDowell and Mansouri in \cite{McD-Mansouri}. 
This action is generalized
to the case of the Moyal ($\theta$-constant) NC space. In the next section we describe
in details the commutative $SO(2,3)$ gravity theory. In Section 3 the NC $SO(2,3)_\star$
gauge theory via the SW map is introduced. As we will see, a problem arises when symmetry breaking
is imposed and we discuss possible solutions. The NC gravity action is presented in Section 4.
It consists of three terms: the Gauss-Bonnet term, the Einstein-Hilbert term and the
cosmological constant term. We calculate the second order corrections for these actions.
Especially, we write the second order corrections in a manifestly gauge covariant way.
In the last section we analyze our results and discuss various applications, remaining
problems and the future work.

\section{AdS gauge theory on commutative spacetime}

The group $SO(2,3)$ is the isometry group of anti-de Sitter space. Anti-de Sitter space is a
maximally symmetric space with negative constant curvature. The $so(2,3)$ algebra or the
AdS algebra consists of $10$ generators, $M_{AB}$. Indices $A,B,\dots$ take values
$0,1,2,3,5$. The generators obey the following commutation relations:
\be
[M_{AB},M_{CD}]=i(\eta_{AD}M_{BC}+\eta_{BC}M_{AD}-\eta_{AC}M_{BD}-\eta_{BD}M_{AC})\ .
\label{AdSalgebra}
\ee
The $5 D$ metric is $\eta_{AB}={\rm diag}(+,-,-,-,+)$.
A representation of this algebra can be constructed starting from the Clifford generators
$\Gamma_{A}$ in $5D$ Minkowski space, which satisfy
\be
\{\Gamma_A,\Gamma_B\}=2\eta_{AB}\ . \nonumber
\ee
Then the generators $M_{AB}$ are
\be
M_{AB}=\frac{i}{4}[\Gamma_A,\Gamma_B]\ .\label{MAB}
\ee
If by $\gamma_a$, $a=0,1,2,3$, we denote the gamma matrices in four dimensional Minkowski space
$M_4$ then the gamma matrices in
$5D$ are $\Gamma_A=(i\gamma_a\gamma_5,\gamma_5)$.
$\gamma_5$ is defined by $\g^5=\g_5=i\g^0\g^1\g^2\g^3.$ It is easy to show that
\bea
M_{ab} &=&\frac{i}{4}[\gamma_a,\gamma_b]=\frac12\sigma_{ab}\ ,\nonumber\\
M_{5a} &=&\frac{1}{2}\gamma_a\ . \label{Maba5}
\eea

If we introduce momenta $P_a=\frac{1}{l}M_{a5}$, where $l$ is a constant with the dimension
of length  the AdS algebra (\ref{AdSalgebra}) becomes
\bea
[M_{ab},M_{cd} ] &=& i(\eta_{ad}M_{bc}+\eta_{bc}M_{ad}-\eta_{ac}M_{bd}-\eta_{bd}M_{ac})
\nonumber\\
{[}M_{ab},P_{c}{]} &=& i(\eta_{bc}P_{a}-\eta_{ac}P_{b})\nonumber\\
{[}P_a,P_b{]} &=&-i\frac{1}{l^2}M_{ab}\ . \label{AdSPoincAlg}
\eea
In the limit $l\to\infty$ the AdS algebra reduces to the usual Poincare algebra in
$M_4$. This is the Wigner-Inon\" u contraction of the AdS algebra.
Useful relations concerning the algebra and the traces of $\gamma$ matrices are given in
Appendix A.

Let us assume that the space-time has the structure of the $4$ dimensional Minkowski space
$M_4$ and follow the usual steps for constructing a gauge theory on $M_4$. To each point
of $M_4$ we attach a tangent space representing a copy of anti-de
Sitter space AdS. The AdS group $SO(2,3)$ acts on matter fields in the tangent space as
a group of internal symmetries. The gauge field takes values in the AdS algebra,
\be
\omega_\m = \frac{1}{2}\omega_\m^{AB}M_{AB}=\frac{1}{4}\omega_\m^{ab}\sigma_{ab}-
\frac12\omega_\m^{a5}\gamma_a .\label{GaugePotAds}
\ee
The gauge potential $\omega_\m^{AB}$ decomposes into
$\omega_\m^{ab}$ and $\omega^{a5}_\m$. The transformation law of the $SO(2,3)$ potential
is given by
\be
\delta_\epsilon\omega_\mu=\pa_\mu\epsilon-i[\omega_\m,\epsilon], \label{TrLawOmegaAB}
\ee
or in components
\bea
\delta_\epsilon\omega_\mu^{AB} &=& \pa_\m\epsilon^{AB}-\epsilon^{A}_{\ C}\omega_\m^{CB}
+\epsilon^{B}_{\ C}\omega_\m^{CA}\ ,\label{DeltaOmegaAB}\\
\delta_\epsilon\omega_\mu^{ab} &=& \pa_\m\epsilon^{ab} -\epsilon^{a}_{\  c}\omega_\m^{cb}
+\epsilon^{b}_{\ c}\omega_\m^{ca} -\epsilon^{a}_{\  5}\omega_\m^{5b}
+\epsilon^{b}_{\ 5}\omega_\m^{5a} \ ,\nn\\
\delta_\epsilon \omega^{a5}_\m &=& \partial_\mu \epsilon^{a5}
-\epsilon^a_{\ c}\omega^{c5}_\m  +\epsilon^5_{\ c}\omega^{ca}_\m \ . \nn
\eea
The field strength is defined in the usual way by
\be
F_{\m\n}=\pa_\m\omega_\n-\pa_\n\omega_\m-i[\omega_\m,\omega_\n]=\frac{1}{2}F^{AB}M_{AB}
\ . \label{FAB}
\ee
Just like the gauge potential, the components of the field strength tensor, $F_{\m\n}^{\ AB}$
can be split into $F_{\m\n}^{\ ab}$ and $F_{\m\n}^{\ a5}$ . It is easy to  show that
\be
F_{\m\n}=\Big( R_{\m\n}^{\ ab}-\frac{1}{l^2}(e_\m^ae_\n^b-e_\m^be_\n^a)\Big)
\frac{\sigma_{ab}}{4} - F_{\m\n}^{\ a5}\frac{\gamma_a}{2}\ , \label{FabFa5}
\ee
where
\bea
R_{\m\n}^{\ ab} &=& \pa_\m\omega_\n^{ab}-\pa_\n\omega_\m^{ab}+\omega_\m^{ac}\omega_\n^{cb}
-\omega_\m^{bc}\omega_\n^{ca} \label{Rab}\\
lF_{\m\n}^{\ a5} &=& D_\m e^a_\n-D_\n e^a_\m = T_{\m\n}^a .\label{Ta}
\eea
Under the local AdS transformation the field strength transforms as
\be
\delta_\epsilon F_{\m\n}=i[\epsilon, F_{\m\n}], \label{TrLawFAB}
\ee
or more explictly
\bea
\delta_\epsilon F^{\ ab}_{\m\n}&=&-\epsilon^{ac} F_{\m\n c}^{\ \ \ b}
+\epsilon^{bc} F_{\m\n c}^{\ \ \  a}-\epsilon^{a5} F_{\m\n 5 }^{\ \ \ b}
+\epsilon^{b5} F_{\m\n5 }^{\ \ \ a} \nn\\
\delta_\epsilon T^{a}_{\m\n}&=&-\epsilon^{ac} T_{\m\n c}+\epsilon^{5c}
F_{\m\n c}^{\ \  \  a}\ .\label{DeltaFabFa5}
\eea
Looking at equations (\ref{GaugePotAds}), (\ref{DeltaOmegaAB}), (\ref{FabFa5}) and
(\ref{DeltaFabFa5}) one is tempted to put $\epsilon^{a5} =0$ and identify $\omega^{ab}_\mu$
with the spin connection of the Poincar\' e gauge theory, $\omega^{a5}_\mu$ with the
vielbeins, $R^{ab}_{\mu\nu}$ with the curvature tensor and $F^{a5}_{\mu\nu}$ with
the torsion.

Indeed, it was shown in the seventies that one can really do such an identification and relate AdS
gauge theory with GR. One way was introduced by MacDowel and Mansouri \cite{McD-Mansouri}.
They start from the $SO(2,3)$ gauge theory but make an additional assumption: that all fields in 
the theory transform covariantly under the
action of infinitesimal diffeomorphisms. The action is written in a way which breaks
the $SO(2,3)$ gauge symmetry down to $SO(1,3)$ and it is invariant under the infinitesimal
diffeomorphisms. Then one can identify $\omega^{a5}_\m$ with the vielbein and after going
to the second order formalism obtain GR\footnote{This holds if there are no spinors
in the theory. If the spinor fields appear, the torsion is nonzero and the pure gravity part of the 
theory does not reduce to GR.}. A similar approach was discussed by
Towsend in \cite{Towsend}.

A more elegant way of relating AdS gauge theory with GR was introduced by Stelle and West
\cite{stelle-west}. They also start from the $SO(2,3)$ gauge theory, but they spontaneously
break it down to $SO(1,3)$. Their starting action is invariant under the full $SO(2,3)$ gauge
symmetry and they introduce one additional auxiliary field in order to perform the symmetry
breaking. In a particular gauge, their action reduces to the MacDowel-Mansouri action which
is invariant under the $SO(1,3)$ gauge symmetry and
again the fields $\omega^{a5}_\m$ can be interpreted as vielbeines. In other gauges the
$SO(2,3)$ symmetry is realized nonlinearly, while the $SO(1,3)$ subgroup is realized linearly.
In that way the diffeomorphism invariance follows from the spontaneous symmetry breaking (SSB)
and does not have to be introduced by hand at the very beginning.

Now we focus on constructing the $SO(2,3)$ gauge invariant action following the Stelle-West
approach. The action invariant under the the $SO(2,3)$ gauge transformations is given by
\be
S = \frac{il}{64\pi G_N}\tr \int{\rm d}^4x \epsilon^{\mu\nu\rho\sigma}
F_{\mu\nu} F_{\rho\sigma}\phi
+ \lambda\int {\rm d}^4x\Big(\frac{1}{4}\tr \phi^2-l^2\Big) \ ,\label{dejstvo1}
\ee
where $G_N$ is the Newton gravitational constant and $\l$ is the Lagrange multiplier and an
additional auxiliary field
$\phi=\phi^A \Gamma_A$ transforming in the adjoint representation of $SO(2,3)$ is introduced
\be
\delta \phi = i[\epsilon,\phi]\ . \label{DeltaPhi}
\ee
The field $\phi$ is constrained by the condition $\phi_A\phi^A=l^2$. Choosing $\phi^a=0,\
\phi^5=l$ the $SO(2,3)$ symmetry is broken spontaneously to $SO(1,3)$ and we  obtain
the action
\bea
S &=& \frac{il^2}{64\pi G_N}\epsilon^{\m\n\r\s}\int {\rm d}^4 x\tr(F_{\m\n}
F_{\r\s}\gamma_5)\nn \\
&=& -\frac{1}{16\pi G_N}\int {\rm d}^4 x\Big[\frac{l^2}{16}\epsilon^{\m\n\r\s}
\epsilon_{abcd}R_{\m\n}^{\ ab}R_{\r\s}^{\ cd} + eR + 2e\Lambda \Big]\ ,
\label{comut.dejstvo,triclana}\eea
where $\Lambda =-3/l^2$ and $e=\det (e_\m^a)$. In the first line we inserted expansions
(\ref{FabFa5}) and (\ref{Rab}) and after some standard manipulation with indices and traces we obtained the
second line. The action (\ref{comut.dejstvo,triclana}) appeared for the first time in the paper by MacDowell and Mansouri 
\cite{McD-Mansouri}.

The vielbeins and spin connection are independent variables and in addition, the spin
connection does not propagate. The last statement follows from the action
(\ref{comut.dejstvo,triclana}). Varying the action with respect
to the spin connection we obtain an equation which relates connection and vielbein. In
this way we can express the spin connection in terms of the vielbein. Since there is no
fermionic matter in the action (\ref{comut.dejstvo,triclana})
this equation gives
vanishing of the torsion.  In that case the first term in (\ref{comut.dejstvo,triclana}) is the
Gauss-Bonnet term; it is a topological term and does not contribute to the equations of
motion. The second term is the Einstein-Hilbert action, while the last term is the
cosmological constant term. From the vielbeins $e_\mu^a$ we can construct the metric tensor
\be
g_{\m\n}=\eta_{ab}e_\m^a e_\n^b\ . \label{MetrTenzor}
\ee
The action (\ref{comut.dejstvo,triclana}) is invariant under the $SO(1,3)$ gauge
transformations. This action is in addition invariant under the infinitesimal
diffeomorphisms as a consequence of spontanious symmetry breaking, as discussed in
\cite{stelle-west}.

\section{NC $SO(2,3)_\star$ gauge theory}

In this section we try to generalize the model (\ref{dejstvo1}) to the NC case. We will not
discuss the Seiberg-Witten map and related calculations in details. Instead we will show that the mechanism of
spontaneous symmetry breaking (SSB) does not work in this case and we will suggest a solution to this problem .

We work with the simplest form of noncommutativity, canonical or
$\theta$-constant noncommutativity,
\begin{equation}
[\hat{x}^\m, \hat{x}^\n] =i \theta^{\m\n},
\end{equation}
with the constant antisymmetric matrix $\theta^{\m\n}$. Following the approach of
deformation quantization we represent noncommutative functions as functions of
commuting coordinates and algebra multiplication with the Moyal-Weyl
$\star$-product (\ref{moyal})
\begin{eqnarray}
\hat{f} (\hat{x}) &\mapsto& f(x) \nonumber\\
\hat{f} (\hat{x})\hat{g} (\hat{x}) &\mapsto& (f\star g)(x) .\nonumber
\end{eqnarray}

In order to construct the NC $SO(2,3)_\star$ gauge theory we use the enveloping algebra approach
and the Seiberg-Witten map developed in \cite{jssw}. Under the infinitesimal
NC $SO(2,3)_\star$ gauge transformations the NC gauge field $\hat{\omega}_\m =
\frac{1}{2}\hat{\omega}_\m^{AB}M_{AB}$ transforms as
\begin{equation}
\delta_\epsilon^\star{\hat\omega}_\m = \pa_\m{\hat\Lambda}_\epsilon
+ i[ {\hat\Lambda}_\epsilon\ds {\hat\omega}_\m] ,\label{SO23Omega}
\end{equation}
with the NC gauge parameter $\hat{\Lambda}_\epsilon$. These transformations close the
algebra
\be
[\delta_{\epsilon_1}^\star\ds\delta_{\epsilon_2}^\star]=\delta_{-i[\epsilon_1, \epsilon_2]}^
\star\ ,\label{SO23GaugeAlgebra}
\ee
provided that the gauge parameter $\hat{\Lambda}_\epsilon$ is in the enveloping algebra of
the $so(2,3)$ algebra. The basic idea of the Seiberg-Witten map is that all noncommutative
variables (gauge parameter, fields) can be expressed in terms of the corresponding
commutative variables as power series in noncommutativity (NC)
parameter $\theta^{\mu\nu}$. In case of the NC gauge parameter the expansion is
\be
\hat{\Lambda}_\epsilon = \Lambda^{(0)} + \Lambda^{(1)} + \dots .\nonumber
\ee
Zeroth order solution is just the commutative gauge parameter $\epsilon =
\frac{1}{2}\epsilon^{AB}M_{AB}$. First order solution (and all higher order solutions)
follows from equation (\ref{SO23GaugeAlgebra}) and it is given by
\be
\hat{\Lambda}^{(1)} = -\frac{1}{4}\theta^{\alpha\beta}\{\omega_\alpha, \partial_\beta\epsilon
\} \ , \label{SO23Lambda1}
\ee
where $\omega_\mu= \frac{1}{2}\omega_\mu^{AB}M_{AB}$ is the commutative gauge potential.
Using the solutions of the SW map given in \cite{kayahan} one can write the
gauge field $\hat{\omega}_\mu$, the field strength tensor $\hat{F}_{\m\n}$ and the field
$\hat{\phi}$ transforming in the adjoint representation in terms of the corresponding
commutative fields $\omega_\m$, $F_{\m\n}$ and $\phi$. Then the action (\ref{dejstvo1}) is
generalized to
\be
S = \frac{il}{64\pi G_N}\tr \int{\rm d}^4x \epsilon^{\mu\nu\rho\sigma}
\hat{F}_{\mu\nu}\star \hat{F}_{\rho\sigma}\star {\hat \phi}
+\lambda\int {\rm d}^4x\Big(\frac{1}{4}\tr \hat{\phi}\star\hat{\phi}-l^2\Big) \ .
\label{NCdejstvo1}
\ee
The field $\hat{\phi}$ is constrained by\footnote{The integral is cyclic
\be
\int {\rm d}^4 x f\star g\star h = \int {\rm d}^4 x h\star f\star g \ .\nonumber
\ee
Especially $\int {\rm d}^4 x f\star g = \int {\rm d}^4 x g\star f =\int {\rm d}^4 x fg$.}
$\hat{\phi}\hat{\phi} =l^2$. Using this constraint to break symmetry from
$SO(2,3)_\star$ to $SO(1,3)_\star$ implies the action
\be
S = \frac{il^2}{64\pi G_N}\tr\int {\rm d}^4 x \epsilon^{\mu\nu\rho\sigma}
{\hat F}_{\mu\nu}\star {\hat F}_{\rho\sigma}\gamma_5
\ .\label{NCdejstvo2}
\ee
This action is supposed to have the NC $SO(1,3)_\star$ gauge symmetry. Unfortunately, it
has not. Therefore, when expanded in orders of NC parameter using the SW map
it will also not have the commutative $SO(1,3)$ symmetry and therefore it will not be possible to
reconstruct GR in the commutative limit. The easiest way to see why the mechanism of SSB fails
in this case is to look at the gauge parameter $\hat{\Lambda}_\epsilon$. If we put
$\epsilon^{a5} =0$ in the solution of the SW map for $\hat{\Lambda}_\epsilon$
(\ref{SO23Lambda1}) we obtain
\bea
\hat{\Lambda}^{(1)}_{so(2,3)} &=& -\frac{1}{4}\theta^{\alpha\beta}\{\omega_\alpha, \partial_\beta\epsilon
\} \nonumber\\
&=& -\frac{1}{4}\theta^{\alpha\beta}\{\frac{1}{4}\omega_\alpha^{ab}\sigma_{ab}
-\frac{1}{2} \omega_\alpha^{a5}\gamma_a, \frac{1}{4}\partial_\beta\epsilon^{cd}
\sigma_{cd} \} \nonumber\\
&=& -\frac{1}{64}\theta^{\alpha\beta}\omega_\alpha^{ab}\partial_\beta\epsilon^{cd}
\{ \sigma_{ab}, \sigma_{cd}\}
+ \frac{1}{32}\theta^{\alpha\beta}\omega_\alpha^{a5}\partial_\beta\epsilon^{cd}
\{ \gamma_a, \sigma_{cd}\} \ .\nonumber
\eea
The second term in the last line produces a term proportional to $\gamma_a\gamma_5$
which is not in the enveloping algebra of $SO(1,3)$. Remember that in this
representation the enveloping algebra of $SO(1,3)$ consists of $\sigma_{ab}$, $I$ and
$\gamma_5$. More explicitly, the SW map for the $SO(1,3)_\star$ gauge symmetry gives
\be
\hat{\Lambda}^{(1)}_{so(1,3)} = -\frac{1}{64}\theta^{\alpha\beta}\omega_\alpha^{ab}\partial_\beta\epsilon^{cd}
\{ \sigma_{ab}, \sigma_{cd}\} \ \nonumber
\ee
and there is no $\gamma_a\gamma_5$-valued term.

A way to solve this problem is to expand the action (\ref{NCdejstvo1}) up
to second order in the NC parameter using the SW map solutions first. The expanded
action has commutative $SO(2,3)$ symmetry which can then be broken to
commutative $SO(1,3)$ symmetry. The obtained action could then be analyzed. This work we
postpone for the next publication.

\section{AdS inspired NC gravity}

We saw that one cannot spontaneously break the NC $SO(2,3)_\star$ gauge symmetry and
obtain GR in the commutative limit. However, one can still choose the MacDowell-Mansouri action
(\ref{comut.dejstvo,triclana}) as the starting
point and work with the NC $SO(1,3)_\star$ gauge symmetry from the very beginning. Then
the SW map guarantees that the expanded action will have commutative $SO(1,3)$ gauge
symmetry. That enables the reconstruction of GR in the commutative limit.

With this motivation, let us write the NC generalization of the action
(\ref{comut.dejstvo,triclana}). It is given by
\bea
S &=& \frac{il^2}{64\pi G_N}\int {\rm d}^4x \epsilon^{\m\n\r\s}
\Big[\tr(\hat{ R}_{\m\n}\star \hat{ R}_{\r\s}\gamma_5)\nn\\
&&-\frac{i}{l^2}\tr(\hat{ R}_{\m\n}\star\hat{ E}_\r\star  \hat{E}_\s\gamma_5)
-\frac{1}{4l^4}\tr({\hat E}_\m \star{\hat E}_\n \star{\hat E}_\r\star
{\hat E}_\s\gamma_5)\Big] \ , \label{ncaction}
\eea
with noncommutative vielbeins ${\hat E}_\m$ and noncommutative curvature 
${\hat R}_{\m\n}$ defined by
\be
{\hat R}_{\m\n}=\pa_\m{\hat \omega}_\n-\pa_\n{\hat \omega}_\m
-i[{\hat\omega}_\m\ds {\hat\omega}_\n]\ ,\label{nckrivina}
\ee
where ${\hat \omega}_\m$ is the noncommutative $SO(1,3)_\star$ gauge potential.

\subsection{Seiberg-Witten map}

Under the deformed gauge transformations the gauge potential and the curvature tensor
transform as
\bea
\delta_\epsilon^\star{\hat\omega}_\m &=& \pa_\m{\hat\Lambda}_\epsilon
-i[ {\hat\omega}_\m\ds{\hat\Lambda}_\epsilon]\nn\\
\delta^\star_\epsilon{\hat R}_{\m\n} &=& i[{\hat\Lambda}_\epsilon\ds{\hat R}_{\m\n}]\ ,
\label{NCTrLawOmegaR}
\eea
where ${\hat\Lambda}_\epsilon$ is the noncommutative gauge parameter.
The NC vielbein $\hat{E}_\mu$ transforms in the adjoint representation
\be
\delta_\epsilon^\star{\hat E}_\m = i[{\hat\Lambda}_\epsilon\ds{\hat E}_\m]\ .
\label{deltaStarE}
\ee
Notice that all noncommutative fields belong to the enveloping algebra of $SO(1,3)$.
For example, the $\star$-commutator in (\ref{deltaStarE}) does not close in the Lie
algebra. This means that in the NC theory we have (apparently) infinitely many new
degrees of freedom compared with the commutative theory.

We have said before that the problem of additional degrees of freedom is solved by
the Seiberg-Witten map. We repeat once again that the basic idea of this map is that
all NC variables can be expressed in terms of the corresponding commutative variables
and their derivatives. These expressions are power series expansions in the NC parameter
\bea
{\hat\Lambda}_{\epsilon}&=&\epsilon+{\hat\L}^{(1)}+{\hat\L}^{(2)}+\dots\ ,\nn\\
\htom_\m &=&\omega_\m+\htom_\m^{(1)}+\htom_\m^{(1)}+\dots\nn\\
\htE_\m &=&e_\m+\htE_\m^{(1)}+ \htE_\m^{(2)}+\dots \ , \nonumber
\eea
where the higher order corrections are functions of the commutative variables
$\epsilon$, $\omega_\m$, $e_\m$ and their derivatives. We will not solve the SW map for our NC fields but use
the results already present in the literature \cite{kayahan}, \cite{PC11} and apply them
to the case of $SO(1,3)_\star$ NC gauge group.

The requirement that the commutator of two NC gauge transformations is a NC
gauge transformation again
\be
[\delta_\a^\star\ds\delta_\b^\star]=\delta_{-i[\a,\b]}^\star \label{SWEqLabda}
\ee
gives the solution for $\Lambda_{\epsilon}^{(1)},
\Lambda_{\epsilon}^{(2)},\dots$. The recursive relation between the $(n+1)$st order and
the $n$th order solution is given by
\be {\hat\L}^{(n+1)} = -\frac{1}{4(n+1)}\theta^{\kappa\lambda}\Big( \{\hat {\omega}_\kappa \ds
\pa_\lambda {\hat \epsilon} \}\Big)^{(n)} \ , \label{RecRelLambda}
\ee
where $(A\star B)^{(n)} = A^{(n)}B^{(0)} + A^{(n-1)}B^{(1)} + \dots
+ A^{(0)}\star ^{(1)} B^{(n-1)} + A^{(1)}\star ^{(1)} B^{(n-2)} +\dots$ includes all
possible terms of order $n$. The explicit expressions for
${\hat \Lambda}^{(1)}_\epsilon$ and ${\hat \Lambda}^{(2)}_\epsilon$ in the case for the
$SO(1,3)_\star$ gauge group are given in Appendix B.
Since $\epsilon$ and $\omega_\m$  contain $\sigma_{ab}$ matrices then the
noncommutative gauge parameter has the following structure
\be {\hat\L}_\epsilon = \frac{1}{4}\Lambda^{ab}\s_{ab} + \L I
+ \L^5\gamma_5\ . \label{SWmapLambdaStructure}
\ee
This means that $[{\hat\L}_\epsilon,\gamma_5]=0. $ Using this fact it is easy to prove
that the action (\ref{ncaction}) is invariant under the NC $SO(2,3)_\star$ gauge group.

Solving the equation
\be
{\hat\omega}_\m(\omega)+\delta_\epsilon^\star{\hat\omega}_\m(\omega)={\hat\omega}_\m(\omega+
\delta_\epsilon\omega) \label{DeltaOmegaSO13}
\ee
order by order in the NC parameter we can express noncommutative gauge
potential $\hat{\omega}_\m$ in terms of the commutative gauge potential $\omega_\mu$.
The recursive solution in this case is given by
\be
{\hat\omega}_\m^{(n+1)}= -\frac{1}{4(n+1)}\theta^{\kappa\lambda}
\Big( \{{\hat \omega}_\kappa \ds \pa_\lambda{\hat \omega}_\m + {\hat R}_{\l\m}\} \Big)^{(n)} .\label{RecRelOmega}
\ee
Looking at the form of this solution, we conclude that the gauge field ${\hat\omega}_\m$ has to be of the form
\be
{\hat\omega}_\m = \frac{1}{4}{\tilde\omega}^{ab}_\m\sigma_{ab} + {\tilde\omega}_\m I + 
{\tilde\omega}^5_\m\gamma_5. \label{SWmapOmegaStructure}
\ee

The solution for the curvature tensor ${\hat R}_{\mu\nu}$ follows from
the definition (\ref{nckrivina}). The recursive formula is
\bea
{\hat R}_{\m\n}^{(n+1)} &=& -\frac{1}{4(n+1)}\theta^{\kappa\lambda}\Big( \{ {\hat \omega}_\kappa \ds
\partial_\lambda {\hat R}_{\mu\nu} + D_\lambda {\hat R}_{\mu\nu} \} \Big)^{(n)} \nn\\
&& +\frac{1}{2(n+1)}\theta^{\kappa\lambda}\Big( \{ {\hat R}_{\mu\kappa}, \ds {\hat R}_{\nu\lambda} \}
\Big)^{(n)} \label{RecRelR}
\eea
and one can check that
\be
{\hat R}_{\mu\nu} = \frac{1}{4}{\tilde R}^{\ ab}_{\m\n}\sigma_{ab} + {\tilde R}_{\m\n} I +
{\tilde  R}^5_{\m\n}\gamma_5. \label{SWmapRStructure}
\ee

The noncommutaive vielbein transforms in the adjoint representation of the NC $SO(1,3)_\star$ gauge group 
\be
\delta_\epsilon^\star{\hat E}_\m = i[{\hat\Lambda}_\epsilon\ds{\hat E}_\m]\  . \label{DeltaESO13}
\ee
The recursive solution is given by
\be
\htE_{\m}^{(n+1)} = -\frac{1}{4(n+1)}\theta^{\kappa\lambda} \Big( \{{\hat \omega}_\kappa \ds \pa_\l {\hat E}_\m + D_\l {\hat E}_\m \} \Big)^{(n)} \label{RecRelE}
\ee
with $D_\l {\hat E}_\m = \partial_\l {\hat E}_\m -i [{\hat \omega}_\l \ds {\hat E}_\m ]$.
The NC vielbein has the structure
\be
{\hat E}_\m = E^a_\m\gamma_a + E^{5a}_\m\gamma_a\gamma_5 \ . \label{SWmapEStructure}
\ee

\subsection{Action}

Using the Sieberg-Witten map solutions (\ref{RecRelOmega}), (\ref{RecRelR}) and (\ref{RecRelE}), 
the action (\ref{ncaction}) becomes a power series expansion
in NC parameter $\theta^{\mu\nu}$
\be
{\hat S} = S + S^{(1)} + S^{(2)}+ \dots \ , \nn
\ee
where $S^{(1)}$ and $S^{(2)}$ are the first and second order corrections respectively.
The zeroth order coincides  with commutative action (\ref{comut.dejstvo,triclana}).
The first order correction is given by
\bea
S^{(1)}&=& \frac{il^2}{64\pi G_N}\epsilon^{\m\n\r\s}\int {\rm d}^4x \Big( 2\tr(\htR^{(1)}_{\m\n}R_{\r\s}\g_5) \nn\\
&& -\frac{i}{l^2}\tr(\htR_{\m\n}^{(1)}e_\r e_\s\gamma_5)
-\frac{i}{l^2}\tr(R_{\m\n}[\htE_{\r}^{(1)}, e_\s] \gamma_5)
+\frac{1}{2l^2}\theta^{\alpha\beta}\tr (R_{\mu\nu}(\partial_\alpha e_\rho)(\partial_\beta e_\sigma)\gamma_5) \nn\\
&& -\frac{1}{2l^4}\tr[e_\m e_\n(2e_\r \htE_{\s}^{(1)}
+ \frac{i}{2}\theta^{\a\b}\pa_\a e_\r\pa_\b e_\s)\gamma_5]\Big) \ .\label{NCdejstvo3clana1Red}
\eea
Inserting the solutions for ${\hat R}_{\m\n}^{(1)}$ and $\htE_{\m}^{(1)}$ and
calculating the traces\footnote{We use the following notation
 $R_{\m\n}=\frac14 R_{\m\n}^{\ \ ab}\sigma_{ab}$ and $e_\m= e_\m^a\g_a.$} in (\ref{NCdejstvo3clana1Red}) one finds that the first order
correction vanishes. In \cite{ChPL} and \cite{PC09} it was shown that all odd order
corrections vanish if the reality of the action is taken into account. Therefore, the
first non-vanishing correction is the second order and we have to calculate
it explicitly.

The second order correction for the action (\ref{NCdejstvo3clana1Red}) is a sum of three terms
\be
S^{(2)}=S^{(2)}_{GB}+S^{(2)}_{EH}+S^{(2)}_{\Lambda}\ . \nn
\ee
The first term is a deformation of the Gauss-Bonnet topological term; the second term is a deformation 
of the Einstein-Hilbert action and the last term is a correction to the cosmological constant term. 
We will analyze them separately.

\subsubsection{Gauss-Bonnet action}

The second order correction to the Gauss-Bonnet action is given by
\bea
S_{GB}^{(2)}&=&\frac{il^2}{64\pi G_N}\epsilon^{\m\n\r\s}\int {\rm d}^4x\Big( 
\tr(\htR_{\m\n}\star\htR_{\r\s}\gamma_5) \Big) ^{(2)}\nn\\
&=&\frac{il^2}{64\pi G_N}\epsilon^{\m\n\r\s}\int {\rm d}^4x\Big( \frac{i}{2}\epsilon_{abcd}R_{\m\n}^{\ ab}
R_{\r\s}^{(2)cd}+8R_{\m\n}^{(1)}
R_{\r\s 5}^{(1)}\Big)\ , \label{GB2nd}
\eea
where we used (\ref{SWmapRStructure}) and the formulas from Appendix A to calculate
the traces. Now one has to insert the explicit solutions for $\htR_{\m\n}^{(1)}$ and
$\htR_{\m\n}^{(2)}$ which are given in Appendix B, formulas (\ref{SWmapR1}) and
(\ref{SWmapR2}) respectively. The calculation is straightforward but very lengthy. It
can be considerably simplified using the following trick: We take the commutative
gauge field $\omega_\m$ to be a constant field. In the final result we reconstruct
the curvature tensor and its covariant derivatives for this specific choice from
\bea
\omega_{\m c}^{a}\omega_\n^{cb}-\omega_{\m c}^{b}\omega_\n^{ca}&\to& R_{\m\n}^{\ ab}\nn\\
\omega_{\m c}^{a}R_{\n\r}^{cb}-\omega_{\m c}^{b}R_{\n\r}^{ca}&\to&(D_\m R_{\n\r})^{ab}\ .
\eea
Inserting $\omega_\mu =const$ into (\ref{GB2nd}) leads to
\bea
S_{GB}^{(2)}&=& -\frac{l^2}{1024
\pi G_N}\theta^{\kappa\l}\theta^{\r\s}\epsilon^{\m\n\a\b}\epsilon_{abcd}\int {\rm d}^4x\Big[R_{\a\b}^{\ cd}
R_{\m\kappa}^{\ ab}R_{\n\r}^{\ mn}R_{\l\s mn}\nn\\
&&-\frac12
R_{\a\b}^{\ cd}
R_{\m\kappa}^{\ ab}R_{\n\l}^{\ mn}R_{\r\s mn} +R_{\a\b}^{\ mn}
R_{\m\kappa mn}R_{\n\r}^{\ ab}R_{\l\s}^{\ cd}\nn\\
&&+R_{\m\r }^{\ mn}
R_{\n\s mn}R_{\a\kappa}^{\ ab}R_{\b\l}^{\ cd}-\frac12R_{\a\kappa}^{\ ab}
R_{\b\l}^{\ cd}R_{\r\s}^{\ mn}R_{\m\n mn}\Big]+ X
\ . \label{GB2nd1}
\eea
In $X$ we collect all terms that are not written in an explicitly covariant way
\bea
X&=& \frac{l^2}{1024\pi G_N}\theta^{\kappa\l}\theta^{\r\s}\epsilon^{\m\n\a\b}\epsilon_{mnpq}
\int {\rm d}^4x\Big[-R_{\a\b}^{\ ab}R_{\m\kappa ab}\omega_\r^{mn}(D_\sigma R_{\n\l})^{pq}\nn\\
&&-\frac12R_{\m\r}^{\ ab}R_{\n\s ab}\omega_\kappa^{mn}(D_\lambda R_{\a\b})^{pq}
-\frac14R_{\r\s e}^{b}R_{\m\n}^{eb}\omega_\kappa^{mn}(D_\lambda R_{\a\b})^{pq}\Big]\ .\eea
This term can be rewritten in the following form
\bea
X &=& \frac{l^2}{1024\pi G_N}\cdot \frac12\theta^{\kappa\l}\epsilon^{\m\n\a\b}\epsilon_{mnpq}\omega_\kappa^{mn}
(D_\lambda R_{\a\b})^{pq}R_{\m\r}^{\ ab}R_{\n\s ab}\nn\\
&& \Big( 
\theta^{\rho\n}\epsilon^{\m\s\a\b}
+\theta^{\rho\m}\epsilon^{\s\a\b\n}+\theta^{\rho\s}\epsilon^{\a\b\n\m}+\theta^{\rho\a}
\epsilon^{\b\n\m\s}\nn
+\theta^{\rho\b}\epsilon^{\n\m\s\a}\Big). \eea
The expression in the bracket is a totally antisymmetric quantity with five indices in
four dimensional Minkowski space and therefore it vanishes.
Finally, the second order correction to the Gauss-Bonnet action, written in a gauge
covariant way is
\bea
S_{GB}^{(2)}&=& -\frac{l^2}{1024
\pi G_N}\theta^{\kappa\l}\theta^{\r\s}\epsilon^{\m\n\a\b}\epsilon_{abcd}\int {\rm d}^4x\Big[
R_{\a\b}^{\ cd}R_{\m\kappa}^{\ ab}R_{\n\r}^{\ mn}R_{\l\s mn}\nn\\
&&-\frac12 R_{\a\b}^{\ cd}
R_{\m\kappa}^{\ ab}R_{\n\l}^{\ mn}R_{\r\s mn}+R_{\a\b}^{\ mn}
R_{\m\kappa mn}R_{\n\r}^{\ ab}R_{\l\s}^{\ cd}\nn\\&&
+R_{\m\r }^{\ mn}R_{\n\s mn}R_{\a\kappa}^{\ ab}R_{\b\l}^{\ cd}
-\frac12R_{\a\kappa}^{\ ab}R_{\b\l}^{\ cd}R_{\r\s}^{\ mn}R_{\m\n mn}\Big]
\ . \label{GB2ndFinal}
\eea
We see that the correction is of fourth order in the curvature. Although it is a
correction of a topological term, the correction itself seems not to be topological
and it would be interesting to see how it modifies the equations of motion.

\subsubsection{Cosmological constant action}

The second order correction to the cosmological constant action is given by
\bea
S^{(2)}_{\Lambda} &=& -\frac{i}{256\pi G_Nl^2}\epsilon^{\m\n\r\s}\int {\rm d}^4x\tr \Big( 
\htE_\m\star\htE_\n\star\htE_\r\star\htE_\s\gamma_5\Big)^{(2)} \nn\\
&=& -\frac{i}{256\pi G_Nl^2}\epsilon^{\m\n\r\s}\int {\rm d}^4x\tr \Big(
4e_\m e_\n e_\r \htE_\s^{(2)} + 4e_\m e_\n \htE_\r^{(1)}\htE_\s^{(1)}\nn\\
&& + 2 e_\m \htE_\n^{(1)} e_\r \htE_\s^{(1)} +i \theta^{\a\b}\{ \partial_\a \htE_\m^{(1)}, \partial_\b e_\n \}e_\r e_\s\nn\\
&& -\frac14\theta^{\a\b}\theta^{\k\l} ( (\partial_\a\partial_\k e_\m)(\partial_\b\partial_\l e_\n)e_\r e_\s
+ (\partial_\a e_\m)(\partial_\b e_\n)(\partial_\k e_\r)(\partial_\l e_\s) )\nn\\
&& +\frac{i}{2}\theta^{\a\b} \{ \partial_\a e_\m(\partial_\b e_\n), [\htE_\r^{(1)} , e_\s]
\} \Big)\g_5 \ . \label{CC2nd1}
\eea
Now we insert the solutions for $\htE_\m^{(1)}$ and $\htE_\m^{(2)}$ given in
Appendix B and calculate the traces. As in the previous case, we simplify the calculation using the trick
of constant fields. This time however, we take only the commutative vielbein $e_\mu$ to
be constant while the commutative gauge field $\omega_\mu$ is arbitrary. This leads to
the action
\bea
S^{(2)}_{\Lambda}&=&  \frac{1}{16\pi G_Nl^2}\theta^{\kappa\lambda}\theta^{\a\b}\int {\rm d}^4x e\Big(
6 e_a^\m E_\m^{(2)a} - 32 (e^\m_a e^\n_b - e^\n_a e^\m_b)
E_{\m 5}^{(1)a}E_{\n5}^{(1)b}\Big)\nn\\
&=& -\frac{1}{256\pi G_Nl^2}\theta^{\kappa\lambda}\theta^{\a\b}\int {\rm d}^4x e\Big(3\pa_\l\omega_\kappa^{cd}\omega_\a^{ac}\omega_\b^{da}+3\pa_\l\omega_\kappa^{cd}\omega_\a^{da}\omega_\b^{ca}\nn\\
&& +2\omega_\kappa^{ab}\omega_\a^{ba}\omega_\lambda^{cd}\omega_\b^{dc}
+8\omega_\kappa^{ab}\omega_\l^{bc}\omega_\a^{cd}\omega_\b^{da}\Big) \ , \label{CC2nd2}
\eea
where in the last line we contracted the $\epsilon$-symbols. Also, the commutative vielbein
$e_\mu$ is used to convert $SO(1,3)$ indices into coordinate indices
$e_{\m}^a X_a=X_\m $. The local index $a,b,\dots$ are raised by the metric $\eta^{ab}$.
The inverse vielbein  $e_a^\m$ is defined by $e_\m^a e^\m_b=\d_b^a$.

This expression can be rewritten in a covariant form
\bea
S^{(2)}_{\Lambda}&=&-\frac{1}{512\pi G_Nl^2}\theta^{\kappa\lambda}\theta^{\a\b}\int
{\rm d}^4x e \Big(6R_{\k\a}^{\ ab}R_{\l\b ab}-3R_{\a\b}^{\ ab}
R_{\k\l ab}\nn\\
&& + 4R_{\a\b}^{\ \ \g\d}(D_\kappa e_\m)^a (D_\l e_\gamma)^b(e_a^\m e_{\d b}
+ e_b^\m e_{\d a})\label{CC2ndFinal} \\
&& -4R_{\a\b}^{\ \ \g\d}(D_\kappa e_\g)^a(D_\l e_{\delta})_a
+ 4(D_\kappa D_\a e_\m )^a (D_\l D_\b e_\n )^b(e_a^\m e_b^\n-e_b^\m e_a^\n)\nn\\
&& -8R_{\k\a}^{\ \ \g\d}(D_\b e_\g)^a(D_\l e_{\delta})_a
+ 8R_{\k\a}^{\ \ \g\d}(D_\b e_\g)^a(D_\l e_{\m})^b(e_a^\m e_{\d b}
+ e_b^\m e_{\d a}\Big) \ . \nn
\eea
The correction has terms that are second, first and zeroth order in the curvature.

\subsubsection{Einstein-Hilbert action}

Finally, we look at the second order correction for the Einstein-Hilbert action. It is
given by
\be
S_{EH}^{(2)} = \frac{1}{64\pi G_N}\epsilon^{\m\n\r\s}\int {\rm d}^4x \tr \Big( {\hat E}_\m\star
\htE_\n\star\htR_{\r\s}\gamma_5 \Big)^{(2)} \ . \label{SEH2nd1}
\ee
Expanding the noncommutative fields and the $\star$-product up to second order in NC
parameter $\theta^{\m\n}$ we obtain
\bea
S_{EH}^{(2)}&=& \frac{1}{64\pi G_N}\epsilon^{\m\n\r\s}\int {\rm d}^4x\tr\Big(
\frac12[e_\m,\e_\n]\htR^{(2)}_{\r\s}\gamma_5+[{\hat E}_\m^{(1)},e_\n ]
{\hat R}_{\r\s}^{(1)}\gamma_5\nn\\
&& +\frac{i}{2}\theta^{\a\b}\{\pa_\a e_\m,\pa_\b e_\n\}\htR^{(1)}_{\r\s}\gamma_5+
[{\hat E}_\m^{(2)},e_\n]R_{\r\s}\gamma_5+\frac12[\htE_\m^{(1)},\htE_\n^{(1)}]R_{\r\s}\gamma_5
\nn\\
&& +\frac{i}{2}\theta^{\a\b}\{\pa_\a e_\m,\pa_\b\htE_\n^{(1)}\}R_{\r\s}\gamma_5
-\frac{1}{16}\theta^{\a\b}\theta^{\g\d}[\pa_\a\pa_\g e_\m,\pa_\b\pa_\d e_\n]R_{\r\s}\gamma_5\Big) \label{SEH2nd2}.
\eea
The next step is to insert the explicit SW map solutions for the fields. This leads to
a very complicated expression which is not written in a gauge covariant way.
The SW map guarantees that the final result has to be covariant under the commutative
$SO(1,3)$ gauge symmetry. We try to apply the same trick as before when calculating the
corrections to the Gauss-Bonnet and the cosmological constant actions. Unfortunately,
we notice that we cannot reconstruct all the covariant terms uniquely. For example, a
term of the form
\be
\epsilon^{\m\n\r\s}\theta^{\a\b}\theta^{\k\l}\int {\rm d}^4x\tr (\omega_\a \omega_\b
\omega_\k \omega_\l \omega_\m \omega_\n e_\r e_\s \gamma_5) \nn
\ee
can be recognized as
\be
\epsilon^{\m\n\r\s}\theta^{\a\b}\theta^{\k\l}\int {\rm d}^4x\tr ( R_{\a\b}R_{\k\l}R_{\m\n}
e_\r e_\s \gamma_5),\nn
\ee
but it can also be seen as one part of 
\be
\epsilon^{\m\n\r\s}\theta^{\a\b}\theta^{\k\l}\int {\rm d}^4x\tr ( (D_\k R_{\a\b})(D_\l R_{\m\n})
e_\r e_\s \gamma_5),\nn
\ee
Therefore, to avoid this ambiguity we use the method developed in \cite{PLM} to calculate the second order
correction for the action (\ref{SEH2nd1}). The main idea of this method is to write
the SW map solutions for composite fields and to simplify calculations in that way.

For example, the $\star$-product of two noncommutative vielbeins is
\be
\htE_\m\star\htE_\n = e_\m e_\n + (\htE_\m\star\htE_\n)^{(1)}
+(\htE_\m\star\htE_\n)^{(2)}+\dots \ . \label{EstarE}
\ee
Using (\ref{DeltaESO13}) one can check that this product transforms in adjoint representation of the gauge group. The first order of (\ref{EstarE})
\be
(\htE_\m\star\htE_\n)^{(1)} = \htE_\m^{(1)}e_\n + e_\m\htE_\n^{(1)}
+\frac{i}{2}\theta^{\a\b}\pa_\a e_\m\pa_\b e_\n \nn
\ee
can be rewritten in the following form
\be
(\htE_\m\star\htE_\n)^{(1)} = -\frac14\theta^{\a\b}\{\omega_\a,\pa_\b(e_\m e_\n)
+D_\b(e_\m e_\n)\}+\frac{i}{2}\theta^{\a\b}(D_\a e_\m)(D_\b e_\n) .\label{EstarE1RedCov}
\ee
The calculation is straightforward. Notice that the first term in (\ref{EstarE1RedCov})
is a solution of the SW map for the field ${\hat \psi} = \htE_\m\star\htE_\n$ in the
adjoint representation, compare with (\ref{RecRelE}). The second term appears because
the field ${\hat \psi} = \htE_\m\star\htE_\n$ is not a fundamental field
but a product of two fundamental fields. Also notice that the second term is written in
terms of covariant derivatives. This will be a big advantage when
we write the action (\ref{SEH2nd1}) in a gauge covariant form.

One can generalize (\ref{EstarE1RedCov}) and write an expression that is valid to all
orders. We will not do that here, for details look at \cite{PLM}.

In the same way we can find the first order term of
${\hat R}_{\m\n}\star\htE_\r\star\htE_\s$. We consider
${\hat R}_{\m\n}\star\htE_\r\star\htE_\s$ as a $\star$-product of the
curvature tensor ${\hat R}_{\m\n}$ and the composite field $\htE_\r\star\htE_\s$. Then
\bea
(\htR_{\m\n}\star\htE_\r\star\htE_\s)^{(1)}&=&\htR_{\m\n}^{(1)}(e_\r e_\s)+R_{\m\n}(\htE_\r\star\htE_\s)^{(1)}+\frac{i}{2}\theta^{\a\b}\pa_\a( R_{\m\n})\pa_\b (e_\r e_\s)\nn\\
&=&-\frac14\theta^{\a\b}\{\omega_\a,\pa_\b(R_{\m\n}e_\r e_\s)+D_\b(R_{\m\n}e_\r e_\s)\} \nn\\
&&+\frac{i}{2}\theta^{\a\b}(D_\a R_{\m\n})D_\b (e_\r e_\s)\nn\\
&& +\frac12\theta^{\a\b}\{R_{\a\m},R_{\b\n}\}e_\r e_\s+\frac{i}{2}\theta^{\a\b}R_{\m\n}(D_\a e_\r)(D_\b e_\s) \ . \label{REE1redCov}
\eea
The first order correction of Einstein-Hilbert action is
\be
S_{EH}^{(1)}=\frac{1}{64\pi G_N}\epsilon^{\m\n\r\s}\int {\rm d}^4x \tr\Big(\htR_{\m\n}\star({\hat E}_\r\star
\htE_\s)\gamma_5\Big)^{(1)}\ .\label{SEH-1}
\ee
Inserting (\ref{REE1redCov}) in (\ref{SEH-1}) and integrating by parts we obtain
\bea
S_{EH}^{(1)}&=& -\frac{1}{256\pi G_N}\epsilon^{\m\n\r\s}\theta^{\a\b}\int {\rm d}^4 x \tr\gamma_5\Big(\{R_{\a\b},R_{\m\n}\}e_\r e_\s\nn\\
&& -2\{R_{\a\m},R_{\b\n}\}e_\r e_\s-2i R_{\m\n}(D_\a e_\r)(D_\b e_\s)\Big) \ . \label{SEH-1Cov}
\eea
This expression is the same as the second line in (\ref{NCdejstvo3clana1Red}), 
but (\ref{SEH-1Cov}) is written in a manifestly gauge covariant way. As we have 
said before, after taking the traces the correction (\ref{SEH-1Cov}) vanishes.

The second order correction of the Einstein-Hilbert action is generated from the first
order correction (\ref{SEH-1Cov}) as
\bea
S_{EH}^{(2)}&=& -\frac{1}{512\pi G_N}\epsilon^{\m\n\r\s}\theta^{\a\b}\int {\rm d}^4x
\tr\gamma_5\Big(\{\htR_{\a\b}\ds\htR_{\m\n}\}\star \htE_\r \star \htE_\s\nn\\
&& -2\{\htR_{\a\m}\ds\htR_{\b\n}\}\star\htE_\r\star \htE_\s
-2i \htR_{\m\n}\star(D_\a \htE_\r)\star(D_\b \htE_\s)\Big)^{(1)}\ . \label{SEH-2}
\eea
Applying
\bea
(\htR_{\a\b}\star\htR_{\m\n})^{(1)}&=&
-\frac14\theta^{\kappa\l}\{\omega_\kappa,\pa_\l(R_{\a\b}R_{\m\n})+D_\l(R_{\a\b}R_{\m\n})\}\nn\\
&& +\frac{i}{2}\theta^{\k\l}(D_\kappa R_{\a\b})(D_{\l}R_{\m\n})+\frac12\theta^{\kappa\l}(\{ R_{\kappa\a},R_{\l\b}\}R_{\m\n}\nn\\
&&+R_{\a\b}\{R_{\kappa\m,R_{\l\n}}\}) \nn
\eea
and
\bea
(D_\a\htE_\r)^{(1)} &=& -\frac14\theta^{\kappa\l}\{\omega_\kappa,\pa_\l(D_\a e_\r)
+D_\l(D_\a e_\r )\}
+ \frac{1}{2}\theta^{\kappa\l}\{ R_{\k\a}, D_\l e_\r \}\nn\\
(D_\a\htE_\r\star D_\b\htE_\s)^{(1)}&=&-\frac14\theta^{\kappa\l}\{\omega_\kappa,
\pa_\l(D_\a e_\r D_\b e_\s)+D_\l(D_\a e_\r D_\b e_\s)\}\nn\\
&& +\frac{i}{2}\theta^{\kappa\l}(D_\k D_\a e_\r)(D_\l D_\b e_\s)\nn\\
&& +\frac12\theta^{\kappa\l}\Big(\{R_{\kappa\a},D_\l e_\r
\}(D_\b e_\s)+(D_\a e_\r)\{R_{\kappa\b},D_\l e_\s\}\Big) \nn
\eea
we obtain
\bea
S_{EH}^{(2)}&=& -\frac{1}{512\pi G_N}\epsilon^{\m\n\r\s}\theta^{\a\b}\theta^{\kappa\l}\int {\rm d}^4x \tr\gamma_5\Big((-\frac14\{R_{\k\l},\{R_{\a\b},R_{\m\n}\}\}\nn\\
&& +\{R_{\k\l},\{R_{\a\m},R_{\b\n}\}\}+\frac12\{R_{\m\n},\{R_{\k\a},R_{\l\b}\}\}
-2\{R_{\a\m},\{R_{\k\b},R_{\l\n}\}\}\nn\\
&& +\frac{i}{2}[D_\kappa R_{\a\b},D_\l R_{\m\n}]-i[D_\kappa R_{\a\m}, D_\l R_{\b\n}])e_\r e_\s\nn\\
&& +i(\{R_{\a\b}, R_{\m\n}\}-2\{R_{\a\m},R_{\b\n}\})
(D_\kappa e_\r)(D_\l e_\s)
-iR_{\m\n}\{D_\a e_\r,\{R_{\kappa\b},D_\l e_\s\}\}\nn\\
&& +R_{\m\n}(D_\kappa D_\a e_\r)( D_\l D_\b e_\s)\Big) \ .\label{SEH-2Cov}
\eea
The next step is to calculate traces. Using the identities for $\gamma$-matrices given
in Appendix A we obtain
\bea
S_{EH}^{(2)}&=& -\frac{1}{512\pi G_N}\theta^{\a\b}\theta^{\kappa\l}\epsilon^{\m\n\r\s}
\epsilon_{abcd}\int d^4x\Big(e_\r^c e_\s^d(-\frac18 R_{\kappa\l}^{\ \ ab}
R_{\a\b}^{\ \ mn}R_{\m\n mn} \nn\\
&&+\frac12 R_{\kappa\l}^{\ \ ab}
R_{\a\m}^{\ \ mn}R_{\b\n mn} + \frac14 R_{\m\n}^{\ \ ab}
R_{\kappa\a}^{\ \ mn}R_{\l\b mn}- R_{\a\m}^{\ \ ab}
R_{\kappa\b}^{\ \ mn}R_{\l\n mn})\nn\\
&& -\frac18R_{\kappa\l\r\s}R_{\a\b}^{\ \ ab}R_{\m\n}^{\ \ cd}
+\frac12R_{\kappa\l\r\s}R_{\a\m}^{\ \ ab}R_{\b\n}^{\ \ cd}\nn\\
&& +\frac14R_{\m\n\r\s}R_{\kappa\a}^{\ \ ab}R_{\l\b}^{\ \ cd}
-R_{\a\m\r\s}R_{\kappa\b}^{\ \ ab}R_{\l\n}^{\ \ cd}\nn\\
&& +e_\r^c e_\s^d(-(D_\kappa R_{\a\b})^{ ma}(D_\l R_{\m\n})^b_{\ m}+2(D_\kappa R_{\a\m})^{ ma}(D_\l R_{\b\n})^b_{\ m})\nn\\
&& +(R_{\a\m}^{\ \ ab}R_{\b\n}^{\ \ cd}-\frac12R_{\a\b}^{\ \ ab}R_{\m\n}^{\ \ cd})(D_\kappa e_\r)^m (D_\l e_\s)_m
\label{SEH-2CovTr} \\
&& +2 R_{\m\n}^{\ \ md} R_{\kappa\beta}^{\ \ ab} (D_\a e_\r)_m (D_\l e_\s)^c
+R_{\m\n}^{\ \ ab}(D_\kappa D_\a e_\r)^c(D_\l D_\b e_\s)^d \Big) \ . \nn
\eea
As in the case of the cosmological constant term we can contract the $\epsilon$-symbols
and simplify (\ref{SEH-2CovTr}). This gives
\bea S_{EH}^{(2)}&=& -\frac{1}{512\pi G_N}\theta^{\a\b}\theta^{\kappa\l}\int d^4xe \Big( \frac12 R_{\kappa\l}^{\ \ \m\n}
R_{\a\b}^{\ \ \g\delta}R_{\m\n\g\delta}\nn\\
&& +R_{\kappa\l\r\s}
(\frac12R_{\a\b}^{\ \ \m\n}R_{\m\n}^{\ \ \r\s}-2R_{\a\b}^{\ \ \m\s}R_\m^{\ \r}+\frac12RR_{\a\b}^{\ \ \r\s} \nn\\
&& +4R_{\b\n}^{\ \ \r\s}R_\a^{\ \n}+4R_\a^{\ \r}R_\b^{\ \s}-4R_{\a\m}^{\ \ \n\r}R_{\b\n}^{\ \ \m\s})\nn\\
&& -2R_{\kappa\l}^{\ \ \m\n}R_{\a\m}^{\ \ \g\delta}R_{\b\n\g\delta}-RR_{\kappa\a}^{\ \ \g\delta}R_{\l\b\gamma\delta}\nn\\
&& -R_{\m\n\r\s}(2R_{\kappa\a}^{\ \ \m\n}R_{\l\b}^{\ \ \r\s}+4R_{\kappa\a}^{\ \ \m\s}R_{\l\b}^{\ \ \n\r}   )\nn\\
&& -4R_\a^{\ \n}R_{\kappa\b}^{\ \ \g\delta}R_{\l\n\g\delta} + 2R_{\a\m\r\s}(2R_{\kappa\b}^{\ \ \m\n}R_{\l\n}^{\ \ \r\s} \nn\\
&&-4R_{\l}^{\ \r}R_{\kappa\b}^{\ \ \m\s}+
4 R_{\kappa\b}^{\ \ \n\r}R_{\l\n}^{\ \ \m\s}+2R_{\kappa\b}^{\ \ \r\s}R_{\l}^{\ \m})\nn\\
&& +4e_b^\m e_c^\n((D_\kappa R_{\a\b})^{mb} (D_\l R_{\m\n})^c_{\ m}-2(D_\kappa R_{\a\m})^{mb} (D_\l R_{\b\n})^c_{\ m})\nn\\
&& +(D_\kappa e_\r)^m (D_\l e_\s)_m(2 R_{\a\b}^{\ \ \m\n} R_{\m\n}^{\ \ \r\s}
+8 R_{\a\b}^{\ \ \m\r} R_{\m}^{\ \s}+2R R_{\a\b}^{\ \ \r\s}\nn\\
&& -8 R_{\a\m}^{\ \ \r\s} R_{\b}^{\ \m}-8 R_{\b}^{\ \r} R_{\a}^{\ \s}
-8R_{\a\m}^{\ \ \n\r} R_{\b\n}^{\ \ \m \s})\nn\\
&& -2(D_\kappa D_\a e_\r)^c(D_\l D_\b e_\s)^d
(R (e_c^\r e_d^\s -e_c^\s e_d^\r) -3R_{\n}^{\ \r}e_{ c}^\n e_d^\s\nn\\
&& +3R_{\n}^{\ \s}e^\n_c e_d^\r
+R_{\n}^{\ \r}e^\s_c e_d^\n-R_{\n}^{\ \s}e^\r_c e_d^\n
+2R_{\m\n}^{\ \ \r\s}e_c^\m e_d^\n) \nn\\
&& +4(D_\lambda e_\s)^a R_{\m\n}^{\ \ bm}(D_\alpha e_\r)_m (
R_{\k\b}^{\ \ \m\n}(e_a^\r e_b^\s - e_a^\s e_b^\r) \nn\\
&& +2R_{\k\b}^{\ \ \m\s}(e_a^\n e_b^\r - e_a^\r e_b^\n)
-2R_{\k\b}^{\ \ \m\r}(e_a^\n e_b^\s - e_a^\s e_b^\n)  
+2R_{\k\b}^{\ \ \r\s}e_a^\m e_b^\n )\Big)
\ . \label{SEH-2CovTrEpsilon}
\eea
The second order correction to the Einstein-Hilbert action is of the 3rd, 2nd and 1st
order in the curvature. Its implications to the equations of motion have to be
investigated carefully. Since the higher powers of curvature 
enter (\ref{SEH-2CovTrEpsilon}), it is obvious that 
the spin connection propagates and it cannot be expressed in terms of vielbeins.

\section{Conclusion}

Starting from the $SO(2,3)$ gravity theory on four dimensional Minkowski
space, we constructed a NC gravity theory. The starting point was the MacDowell-Mansouri action
(\ref{comut.dejstvo,triclana}) obtained after the spontaneous symmetry breaking of
$SO(2,3)$ to $SO(1,3)$. The construction is done as an expansion in the
NC parameter, using the SW map. As expected, the first order correction vanishes and
for the second order correction one obtains a very complicated expression which not 
written in an explicitly gauge
covariant way. Using the trick of constant fields (connection and vielbein) and the SW
map for composite fields we managed to write the action in a manifestly gauge covariant
way, see (\ref{GB2ndFinal}),
(\ref{SEH-2CovTrEpsilon}) and (\ref{CC2ndFinal}). Let us discuss some of the properties 
of our action. 

The NC action (\ref{GB2ndFinal}),
(\ref{SEH-2CovTrEpsilon}) and (\ref{CC2ndFinal}) is invariant under the 
commutative $SO(1,3)$ gauge
symmetry. However, the question of diffeomorphism invariance remains to be understood
better. Since $\theta^{\m\n}$ is constant matrix, the symmetry under the
usual (commutative, undeformed) diffeomorphisms is broken. Let us briefly discuss the invariance
under the canonically twisted diffeomorphisms introduced in \cite{defgt}. For this purpose we rewrite 
the action (\ref{ncaction}) in the language of forms and use the notation introduced in \cite{defgt}. 
The action is given by
\begin{equation}
S = \frac{il^2}{64\pi G_N}\int \tr \Big( \hat{ R}\wedge_\star \hat{ R}\gamma_5
-\frac{i}{l^2}\hat{ R}\wedge_\star\hat{ E}\wedge_\star  \hat{E}\gamma_5
-\frac{1}{4l^4}{\hat E}\wedge_\star{\hat E}\wedge_\star{\hat E}\wedge_\star
{\hat E}\gamma_5\Big) \ . \label{concl1} 
\end{equation}
The action of infinitesimal twisted
diffeomorphisms is given by the action of the $\star$-Lie derivative along the vector field $\xi=\xi^\mu\partial_\mu$. 
Let us examine how the $\star$-Lie derivative acts on the first term in (\ref{concl1})
\begin{eqnarray}
&& {\cal L}^\star_\xi \int \tr \hat{ R}\wedge_\star \hat{ R}\gamma_5  \nonumber\\
&& = \int \tr \Big( {\rm d} \langle \xi ,\hat{ R}\wedge_\star \hat{ R} \rangle_\star 
+ \langle \xi , {\rm d} (\hat{ R}\wedge_\star \hat{ R}) \rangle_\star
\Big)\gamma_5 \nonumber\\
&& = {\mbox{ surface term} }=0 \ . \nonumber
\end{eqnarray}
In the second line, the second term is an exterior derivative of a $4$-form in $4$ dimension and therefore vanishes, while 
the first term is a surface term. One can perform the similar analysis for the other two terms in (\ref{concl1}), the result 
is the same. Therefore, we conclude that the unexpanded action (\ref{ncaction}) is invariant under the twisted diffeomorphism
symmetry. Note that it is difficult to perform this analysis with the expanded action, one of the problems being that the 
the twisted comultiplication (twisted Leibniz rule) is only defined for the $\star$-product of fields and not for the 
pointwise products. However, from the previous analysis we can conclude that the expanded action is also invariant under 
the twisted diffeomorphism symmetry.

Note that one can also look at the full $SO(2,3)_\star$ NC
action (\ref{NCdejstvo1}). Using the SW map for the composite fields one can write
the second order correction for this action. The expanded action has the
commutative $SO(2,3)$ gauge symmetry. This symmetry can be spontaneously broken 
to the commutative $SO(1,3)$
symmetry. The zeroth order of the action after SSB has to be given by
(\ref{comut.dejstvo,triclana}) but it would be interesting to compare the higher order
corrections with the results obtained here.

In the commutative limit, $\theta^{\m\n}\to 0$, the action (\ref{GB2ndFinal}), 
(\ref{SEH-2CovTrEpsilon}) and (\ref{CC2ndFinal}) reduces to the Einstein-Hilbert action
plus the Gauss-Bonnet and the cosmological constant terms. The first order correction
vanishes, as expected. In the second order correction terms proportional to the 4th
and lower powers of the curvature tensor appear. Although the contractions of the
curvature tensor (Ricci tensor and scalar curvature) appear in (\ref{GB2ndFinal}), 
(\ref{SEH-2CovTrEpsilon}) and (\ref{CC2ndFinal}), note that in most of the terms the 
indices of the curvature tensor are contracted with the indices of the NC parameter
$\theta^{\m\n}$. Therefore, our result does not resemble to $f(R)$ theories. Instead,
it sems to describe a more complicated theory.

Concerning the future research, there are many different directions. For example,
the action (\ref{GB2ndFinal}), (\ref{SEH-2CovTrEpsilon}) and (\ref{CC2ndFinal}) 
can be used to calculate
corrections to classical solutions such as black holes and cosmological solutions. 
It would be interesting to see how the deformation modifies the Hawking temperature
of black hole radiation. In addition, one can investigate quantum behavior of 
the deformed theory and learn how the deformation influences renormalizability.

As we have said before, our NC action contains higher powers of the curvature tensor 
and its covariant derivatives.
These terms become increasingly important as energies become higher. The prime
candidates for the study of these terms are therefore effects in the early universe,
where the epoch of inflation is especially interesting. Higher powers of the
curvature tensor and its contractions have been early identified as possible sources of
inflationary expansion in the early universe. The model of Starobinsky  \cite{star} is
probably the best known representative of the class of models in which the inflationary
expansion is modeled in terms of the higher order terms in the action. It is reasonable
to expect that for some values of $\theta$ inflationary dynamics should be realized as
a result of terms present in the NC action (\ref{GB2ndFinal}),
(\ref{SEH-2CovTrEpsilon}) and (\ref{CC2ndFinal}).
The question of the existence and the
properties of inflationary solutions in the NC action (\ref{GB2ndFinal}),
(\ref{SEH-2CovTrEpsilon}) and (\ref{CC2ndFinal}) requires a dedicated analysis
and it will be presented elsewhere.

One frequently wonders how much physical relevance is there in theories defined on the 
Moyal plane, i.e. with constant noncommutativity parameter. We try to sketch a 
justification for the use of constant noncommutativity using the physics of inflationary
universe. If we assume that the mechanism determining the value of $\theta$ in the very early (preinflationary) 
universe is local, i.e. that the value of $\theta$ acquires values locally or has constant 
components in small local domains, then depending on the particular values of 
$\theta$ components, in some of these domains the action terms (4.66) and (4.77) will lead to inflation 
and in some it will not. The domains with the inflationary dynamics will expand much more than those 
that do not inflate and their value of $\theta$ will be associated to very large 
physical volumes after the inflation. In the parts of the universe which originate from the same 
preinflationary domain it is therefore reasonable to expect that after the inflation 
the components of the $\theta$ will be constant. Although this argument puts additional 
weight on theories of noncommutative spaces with constant $\theta$, a precautionary 
remark is in order. The action  (\ref{GB2ndFinal}),
(\ref{SEH-2CovTrEpsilon}) and (\ref{CC2ndFinal}) was derived using the assumption of 
constant $\theta$. If it is to 
be valid in the preinflationary epoch, the domains in which $\theta$ is constant before 
the inflation should be sufficiently large so that IR corrections to (\ref{GB2ndFinal}),
(\ref{SEH-2CovTrEpsilon}) and (\ref{CC2ndFinal}), 
associated to the size of the domain, are negligible. A more 
precise formulation of the claim on constancy of $\theta$ would be that the domains in 
postinflationary universe with constant $\theta$ are much bigger that the 
domains of constancy of $\theta$ in the preinflationary universe.   

Concerning the experimental constraints on the NC parameter $\theta^{\m\n}$ one can say the following:
The existing studies of inflation in modified theories of gravity (such as $f(R)$
theories and other higher-derivative theories) may not be immediately
applied to constrain $\theta$ from the action (\ref{GB2ndFinal}), (\ref{SEH-2CovTrEpsilon})
and (\ref{CC2ndFinal}). We have said before that in these expressions the $\theta$ 
components are contracted
with the curvature tensor and
its covariant derivatives and $f(R)$ and similar theories do not include such terms.
This property is a distinguishing novelty which
should be useful in the experimental verification/falsification of the theory presented
in this paper.

\vskip1cm \noindent
{\bf Acknowledgement}
\hskip0.3cm
We would like to thank Milutin Blagojevi\' c and Maja Buri\' c for
fruitful discussion and useful comments. The work of M.~D. and V.~R. is supported by project
171031 of the Serbian Ministry of Education and Science. The work of H.~S is supported
by Ministry of Education, Science and Sport of the republic of Croatia under
contract No 098-0982930-2684. This paper is done if the framework of the
Serbian-Croatian bilateral project "Modified gravity theories and the accelerated
expansion of Universe".

\appendix

\section{AdS algebra and the $\gamma$-matrices}

Algebra relations\footnote{$\epsilon^{01235}=+1,\ \epsilon^{0123}=1$}:
\bea \{M_{AB},\Gamma_C\}&=&i\epsilon_{ABCDE}M^{DE}\nn\\
\{M_{AB},M_{CD}\}&=&\frac{i}{2}\epsilon_{ABCDE}\Gamma^{E}+\frac12(\eta_{AC}\eta_{BD}-\eta_{AD}\eta_{BC})\nn\\
{[}M_{AB},\Gamma_C{]}&=&i(\eta_{BC}\Gamma_A-\eta_{AC}\Gamma_B)\nn\\
\Gamma_A^\dagger&=&-\gamma_0\Gamma_A\gamma_0\nn\\
M_{AB}^\dagger&=&\gamma_0M_{AB}\gamma_0\nn\\
\{\sigma_{ab},\sigma_{cd}\}&=&2(\eta_{ac}\eta_{bd}-\eta_{ad}\eta_{bc}+i\epsilon_{abcd}\gamma_5)\nn\\
{[}\sigma_{ab},\gamma_c{]}&=&2i(\eta_{bc}\gamma_a-\eta_{ac}\gamma_b)\nn\\
\{\sigma_{ab},\gamma_c\}&=&2\epsilon_{abcd}\g^5\g^d
\eea
Identities with traces:
\bea
&&\tr (\Gamma_A\Gamma_B)=4\eta_{AB}\nn\\
&&\tr (\Gamma_A)=\tr (\Gamma_A\Gamma_B\Gamma_C)=0\nn\\
&&\tr (\Gamma_A\Gamma_B\Gamma_C\Gamma_D)=4(\eta_{AB}\eta_{CD}-\eta_{AC}\eta_{BD}+\eta_{AD}\eta_{CB})\nn\\
&&\tr (\Gamma_A\Gamma_B\Gamma_C\Gamma_D\Gamma_E)=-4i\epsilon_{ABCDE}\nn\\
&&\tr (M_{AB}M_{CD}\Gamma_E)=i\epsilon_{ABCDE}\nn\\
&&\tr (M_{AB}M_{CD})=-\eta_{AD}\eta_{CB}+\eta_{AC}\eta_{BD}
\eea

\section{SW map for $SO(1,3)_\star$ gauge group}

Here we list the explicit 1st and 2nd order solutions for the gauge parameter
${\hat \Lambda}_\epsilon$,
the gauge field ${\hat \omega}_\mu$, the curvature tensor ${\hat R}_{\mu\nu}$ and
the vielbein ${\hat E}_\mu$ for the $SO(1,3)_\star$ gauge group.
\begin{itemize}
\item Gauge parameter
\bea
{\hat\L}^{(1)} &=& -\frac{1}{4}\theta^{\kappa\lambda}\{\omega_\kappa,
\partial_\lambda\epsilon \} \ , \label{SWmapLambda1}\\
{\hat\L}^{(2)} &=& \Lambda_\epsilon^{ab(2)}\frac{\sigma_{ab}}{4}\nn\\
&=& \frac{1}{32}\theta^{\kappa\lambda}\theta^{\r\s}
\Big(\{\omega_\r,\{\pa_\s\omega_\kappa,\pa_\l\epsilon\}\}
+\{\omega_\r,\{\omega_\kappa,\pa_\s\pa_\l\epsilon\}\}\nn\\
&& +\{\{\omega_\r,\pa_\s\omega_\kappa\},\pa_\l\epsilon\}
-\{\{R_{\r\kappa},\omega_\s\},\pa_\l\epsilon\}\nn\\
&& -2i[\pa_\r \omega_\kappa,\pa_\s\pa_\lambda\epsilon]\Big)\ . \label{SWmapLambda2}
\eea

\item Gauge field
\bea {\hat\omega}_\m^{(1)}&=&-\frac{1}{4}\theta^{\kappa\lambda}
\{\omega_\kappa,\pa_\lambda\omega_\m+R_{\l\m}\}\nn\\
&=& \omega^{(1)}_\m I + \omega_{\m 5}^{(1)}\gamma_5\ , \label{SWmapOmega1}
\eea
with
\bea
\omega^{(1)}_\m &=& -\frac{1}{16}\theta^{\kappa\l} \omega_\kappa^{ab}(\pa_\l\omega_{\m ab} + R_{\l\m ab}), \nn\\
\omega_{\m 5}^{(1)} &=& -\frac{i}{32}\theta^{\kappa\l} \epsilon_{abcd}\omega_\kappa^{ab}(\pa_\l\omega_{\m}^{cd}+R_{\l\m}^{\ cd})\ . \nn
\eea
\bea
{\hat\omega}^{(2)}_\m &=& \frac{1}{4}\omega_\m^{ab(2)}\sigma_{ab}\nn\\ 
&=&-\frac{1}{8}\theta^{\kappa\lambda}
\{{\hat\omega}_\kappa^{(1)},\pa_\lambda\omega_\m+R_{\l\m}\}+
\{{\omega}_\kappa,\pa_\lambda{\hat\omega}_\m^{(1)}+{\htR}_{\l\m}^{(1)}\}\nn\\
&& -\frac{i}{16}\theta^{\kappa\l}\theta^{\a\b}
[\pa_\a\omega_\kappa,\pa_\n(\pa_\l\omega_{\m}+R_{\l\m})]\ , \label{SWmapOmega2}
\eea
where $ \htR_{\m\n}^{(1)}$ is the first order corrections to the curvature tensor (see below).

\item Curvature tensor
\bea
{\hat R}_{\m\n}^{(1)} &=& -\frac{1}{4}\theta^{\kappa\lambda}
\{\omega_\kappa,\pa_\lambda R_{\m\n} + D_\lambda R_{\m\n}\} + \frac{1}{2}\theta^{\kappa\lambda} \{R_{\mu\kappa}, R_{\nu\lambda} \} \nonumber\\
&=& R_{\m\n}^{(1)}I + R_{\m\n5}^{(1)}\gamma^5\ , \label{SWmapR1}
\eea
where
\bea
R_{\m\n}^{(1)} &=& \frac{1}{8} \theta^{\r\s}\Big(R_{\m\r}^{\ ab}R_{\n\s ab}-\omega_\r^{ab}\pa_\s R_{\m\n ab}
-\frac12\omega_{\r ab}(\omega_{\s e}^{a}R_{\m\n}^{\ eb}-\omega_{\s e}^{b}R_{\m\n}^{\ ea})\Big)  \ , \nonumber\\
R_{\m\n5}^{(1)} &=& \frac{i}{16}\theta^{\r\s}\epsilon_{abcd}
\Big(R_{\m\r}^{\ ab}R_{\n\s}^{\ cd}-\omega_\r^{ab}\pa_\s R_{\m\n}^{\ cd}
-\omega_\r^{ab}\omega_{\s e}^{c}R_{\m\n}^{\ ed}\Big)\ . \nonumber
\eea
\bea
{\htR}_{\m\n}^{(2)} &=& \frac{1}{4}R_{\m\n}^{ab(2)}\sigma_{ab}\nn\\
&=&-\frac18\theta^{\kappa\l}\Big({\{}\omega_\kappa,\pa_\l {\htR}^{(1)}_{\m\n}
+(D_\l {\hat R}_{\m\n})^1{\}}+
\{{\htom}_\kappa^{(1)},\pa_\l R_{\m\n}+(D_\l R_{\m\n}){\}}\nn\\
&&-2{\{}R_{\m\kappa},{\htR}_{\n\l}^{(1)}{\}}-2{\{}{\htR}_{\m\kappa}^{(1)}, R_{\n\l}{\}}
\Big) \label{SWmapR2}\\
&& -\frac{i}{16}\theta^{\kappa\l}\theta^{\r\s}\Big( [\pa_\r \omega_\kappa,\pa_\s(\pa_\l R_{\m\n}+D_{\lambda}R_{\m\n})]-2[\pa_\r R_{\m\kappa},\pa_\s R_{\n\l}]\Big)\ , \nn
\eea
where
\bea
{ R}_{\m\n}^{(2)ab}&=&-\frac{1}{8}\theta^{\kappa\lambda}\omega_\kappa^{ab}(4\pa_\l R_{\m\n}^{(1)}
+\frac14\theta^{\a\b}\pa_\a\omega_\l^{cd}\pa_\b R_{\m\n cd})\nn\\
&& -\frac{i}{8}\epsilon^{abpq}\omega_\kappa^{pq}(2\pa_\l R_{\m\n5}^{(1)}
+\frac{i}{16}\theta^{\a\b}\epsilon_{cdef}\pa_\a\omega^{cd}_\l
\pa_\b R^{\ ef}_{\m\n})\nn\\
&& -\frac14\theta^{\kappa\l}\omega^{(1)}_\kappa(\pa_\l R_{\m\n}^{\ ab}+ (D_\l R_{\m\n})^{ab})\nn\\
&& - \frac{i}{8}\theta^{\kappa\l}\epsilon_{abcd}\omega^{(1)}_{\kappa5}(\pa_\l R_{\m\n}^{\ cd}+ (D_\l R_{\m\n})^{cd})\nn\\
&& + \frac12\theta^{\kappa\l}(2R_{\m\kappa}^{\ ab}R_{\n\l}^{(1)}+i\epsilon_{abcd}R_{\m\kappa}^{\ cd}R_{\n\l5}^{(1)})\nn\\
&& +\frac{1}{16}\theta^{\kappa\l}\theta^{\r\s}\Big( \pa_\r\omega_{\kappa e}^a\pa_\s(\pa_\l R_{\m\n}^{\ eb}
+(D_\l R_{\m\n})^{eb})\nn\\
&& -\pa_\r\omega_{\kappa e}^{b}\pa_\s(\pa_\l R_{\m\n}^{\ ea}+(D_\l R_{\m\n})^{ea})\nn\\
&& -2\pa_\r R_{\m\kappa e}^{a}\pa_\s R_{\n\l}^{\ eb} + 2\pa_\r R_{\m\kappa e}^{b}\pa_\s R_{\n\l}^{\ ea}\Big)
\ . \nonumber
\eea

\item Vielbein
\bea
\htE_{\m}^{(1)} &=& -\frac{1}{4}\theta^{\kappa\lambda}
\{\omega_\kappa,\pa_\lambda e_\m + D_\lambda e_\m \} \nn\\
&=& E_{\m d5}^{(1)}\gamma_5\gamma^d\nn\\
&=&-\frac18\theta^{\kappa\lambda}\epsilon_{abcd}\omega_\kappa^{ab}(2\pa_\l e_\m^c+\omega_\l^{ce} e_{\m e})\g_5\g^d \ . \label{SWmapE1}
\eea
\bea \htE_{\m}^{(2)}&=& E_{\m a}^{(2)}\gamma^a\nn\\
&=&-\frac14\theta^{\kappa\lambda}\omega_\kappa^{(1)}( 2\pa_\l e_\m^a+\omega_\l^{ab}e_{\m b})\gamma_a\nn\\
&& -\frac{1}{16}\epsilon_{cdfa}\theta^{\k\l}\omega_{\kappa}^{cd}(2\pa_\l E_{\m5}^{(1)f}+\omega_{\l}^{fe}E_{\m e5}^{(1)}-2i\omega_{\l5}^{(1)}e_\m^f\nn\\
&& +\frac14\theta^{\a\b}\epsilon_{mne}^{\ \ \ \ f}\pa_\a\omega_\l^{mn}\pa_\b e_\m^e)\gamma^a\nn\\
&& +\frac{1}{16}\theta^{\kappa\l}\theta^{\r\s}\pa_\r\omega_\kappa^{ab}
\pa_\s(\pa_\l e_\m^b+\omega_\l^{bd}e_{\m d})\gamma_a \ . \label{SWmapE2}
\eea

\end{itemize}


\begin{thebibliography}{99}

\bibitem{NCknjige}
A.~Connes, {\it Non-commutative Geometry}, Academic Press (1994).

J. Madore, {\it An Introduction to Noncommutative Differential
Geometry and its Physical Applications}, 2nd Edition, Cambridge
Univ. Press (1999).

P.~Aschieri, M.~Dimitrijevi\' c, P.~Kulish, F.~Lizzi and J.~Wess
{\it Noncommutative spacetimes:
Symmetries in noncommutative geometry and field theory},
Lecture notes in physics {\bf 774}, Springer (2009).

\bibitem{DefQuant}
F.~Bayen, M.~Flato, C.~Fronsdal, A.~Lichnerowicz and D.~Sternheimer,
{\it Deformation theory and quantization},
Ann.\ Phys. {\bf 111}, 61 (1978).

D.~Sternheimer,
{\it Deformation quantization: Twenty years after},
AIP Conf.\ Proc.\  {\bf 453}, 107 (1998)
[math.qa/9809056].

Maxim Kontsevich, {\it Deformation quantization of Poisson manifolds, I},
Lett.\ Math.\ Phys.\ {\bf 66}, 157 (2003) [q-alg/9709040].

\bibitem{Seiberg:1999vs}
N.~Seiberg and E.~Witten,
{\it String theory and noncommutative geometry},
JHEP {\bf 09}, 032 (1999) [hep-th/9908142].

\bibitem{jssw}
B. Jur\v{c}o, S.~Schraml, P.~Schupp and J.~Wess,
{\it Enveloping algebra valued gauge transformations for 
non-abelian gauge groups on non-commutative spaces }, Eur.\ Phys.\ J.\ C{\bf 17},
521 (2000) [hep-th/0006246].

B. Jur\v{c}o, L. M\"oller, S.~Schraml, P.~Schupp and J.~Wess,
{\it Construction of non-Abelian gauge theories on noncommutative spaces},
Eur.\ Phys.\ J.\ C{\bf 21}, 383 (2001) [hep-th/0104153].

\bibitem{Buric:2005xe}
M.~Buri\'c, D.~Latas and V.~Radovanovi\'c,
{\it Renormalizability of noncommutative SU(N) gauge theory}, JHEP {\bf 0602}, 046 
(2006) [hep-th/0510133].

M.~Buri\'c, V.~Radovanovi\'c and J.~Trampeti\'c,
{\it The one-loop renormalization of the gauge sector in the noncommutative
standard model}, JHEP {\bf 0703}, 030 (2007) [hep-th/0609073].

\bibitem{martin}
C. P. Martin, {\it The gauge anomaly and the Seiberg-Witten map}, Nucl.\ Phys.\ B {\bf 652}, 72
(2003) [arXiv:hep-th/0211164].

R. Banerjee and S. Ghosh, {\it Seiberg-Witten map and the axial anomaly in 
noncommutative field theory}, Phys.\ Lett.\ B {\bf 533}, 162 (2002), [hep-th/0110177].

R. Banerjee and K. Kumar, {\it Seiberg-Witten maps and commutator anomalies in noncommutative
electrodynamics}, Phys.\ Rev.\  D {\bf 72}, 085012 (2005), [hep-th/0505245].

\bibitem{cjsww}
X.~Calmet, B.~Jur\v co, P.~Schupp, J.~Wess and M.~Wohlgenannt,
{\it The Standard Model on noncommutative spacetime},
Eur.\ Phys.\ J. {\bf C23}, 363 (2002)
[hep-ph/0111115].

P.~Aschieri, B.~Jur\v co, P.~Schupp and J.~Wess,
{\it Noncommutative GUTs, standard model and C, P, T},
Nucl.\ Phys.\ B {\bf 651}, 45 (2003)
[arXiv:hep-th/0205214]

\bibitem{tramp}
W.~Behr, N.~G.~Deshpande, G.~Duplan\v ci\' c, P.~Schupp, J.~Trampeti\' c and J.~Wess,
{\it The Z $\to$ gamma gamma, gg Decays in the Noncommutative Standard Model},
Eur.\ Phys.\ J. {\bf C29}, 441 (2003) [hep-ph/0202121].

B.~Meli\' c, K.~Passek-Kumeri\v cki, P.~Schupp, J.~Trampeti\' c and M.~Wohlgennant,
{\it The Standard Model on Non-Commutative Space-Time: Electroweak
Currents and Higgs Sector}, Eur.\ Phys.\ J. {\bf C42}, 483 (2005)
[hep-ph/0502249].

B.~Meli\' c, K.~Passek-Kumeri\v cki, P.~Schupp, J.~Trampeti\' c and M.~Wohlgennant,
{\it The Standard Model on Non-Commutative Space-ime: Strong Interactions
Included}, Eur.\ Phys.\ J. {\bf C42}, 499 (2005)
[hep-ph/0503064].

\bibitem{defgt}
P.~Aschieri, C.~Blohmann, M.~Dimitrijevi\' c, F.~Meyer, P.~Schupp
and J.~Wess, {\it A Gravity Theory on Noncommutative Spaces},
Class.\ Quant.\ Grav. {\bf 22}, 3511 (2005) [hep-th/0504183 ].

P.~Aschieri, M. Dimitrijevi\' c, F.~Meyer and  J.~Wess, 
{\it Noncommutative Geometry and Gravity}, Class.\ Quant.\ Grav. {\bf
23}, 1883 (2006) [hep-th/0510059].

\bibitem{MilutinKnjiga}
M. Blagojevi\'c, {\it Gravitation and Gauge Symmetries}, Institute of Physics Publication, Bristol (2002).

\bibitem{ChPL} A. H. Chamseeddine, {\it Deforming Einstein's gravity}, 
Phys.\ Lett.\ B {\bf 504} 33 (2001) [hep-th/0009153].

\bibitem{ChPrd04} A. H. Chamseeddine, {\it $SL(2,C)$ gravity with a complex vierbein 
and its noncommutative extension}, Phys.\ Rev.\ D {\bf 69}, 024015 (2004) [hep-th/0309166].

\bibitem{dZanon03} M. A. Cardella and D. Zanon,  
{\it Noncommutative deformation of four-dimensional gravity}, {\bf 20}, L95 (2003) [hep-th/0212071].

\bibitem{BMS-07} R. Banerjee, P. Mukherjee and S. Samanta, {\it Lie algebraic noncommutative gravity},
 Phys. Rev D {\bf 75}, 125020 (2007)
[hep-th/0703128].

\bibitem{MXZ-11} Yang-Gang Miao, Zhao Xue and Shao-Jun Zhang, {\it $U(2,2$ 
gravity on noncommutative space with sympletic structure}, Phys. Rev. D {\bf 83}, 
024023 (2011) [arXiv:1006.4074].
    
\bibitem{PC09} P. Aschieri and L. Castellani, {\it Noncommutative $D=4$ gravity 
coupled to fermions} JHEP, {\bf 0906}, 086 (2009) [arXiv:0902.3823].

\bibitem{PC12} P. Aschieri and L. Castellani, {\it Noncommutative gauge fields 
coupled to noncommutative gravity}, [arXiv:1205.1911]

\bibitem{Mukherjee2} P. Mukherjee and A. Saha, {\it A Note on the noncommutative correction to gravity}, 
Phys.\ Rev D {\bf 74}, 027702 (2006) [hep-th/0605287].

\bibitem{stelle-west} K. S. Stelle and P. C. West, {\it Spontaneously 
broken de Sitter symmetry and the gravitational holonomy group}, Phys.\ Rev D 
{\bf 21}, 1466 (1980).

\bibitem{McD-Mansouri} S. W. MacDowell and F. Mansouri, 
{\it Unified geometrical theory of gravity and supergravity}, Phys.\ Rev.\ Lett. {\bf 38}, 739 (1977).

\bibitem{Towsend} P. K. Towsend, {\it Small-scale structure of spacetime as 
the origin of the gravitation constant}, Phys.\ Rev.\ D {\bf 15},  2795 (1977).


\bibitem{kayahan} K. Ulker and B. Yapiskan, {\it Seiberg-Witten maps to all orders}, 
Phys.\ Rev.\ D {\bf 77}, 065006 (2008) [arXiv: 0712.0506].


\bibitem{PC11} P. Aschieri and L. Castellani, {\it Noncommutative gravity 
coupled to fermions: second order expansion via Seiberg-Witten map}, [arXiv:1111.4822].


\bibitem{PLM} P. Aschieri, L. Castellani and M. Dimitrijevi\'c, {\it Noncommutative gravity at second order 
via Seiberg-Witten map}, [arXiv:1207.4346].

\bibitem{star} A. A. Starobinsky, {\it A New Type of Isotropic Cosmological Models Without Singularity}, 
Phys.\ Lett.\ B {\bf 91}, 99 (1980).





\end{thebibliography}
\end{document}